\documentstyle{article}

\font\tenrm=cmr10
\font\tenit=cmti10
\font\elevenbf=cmbx10 scaled\magstep 1
\font\elevenrm=cmr10 scaled\magstep 1
\font\elevenit=cmti10 scaled\magstep 1

\font\ninerm=cmr9

\renewenvironment{thebibliography}[1]
 { \elevenrm
   \begin{list}{\arabic{enumi}.}
    {\usecounter{enumi} \setlength{\parsep}{0pt}
     \setlength{\itemsep}{3pt} \settowidth{\labelwidth}{#1.}
     \sloppy
    }}{\end{list}}

\begin{document}
\begin{center}
\vglue 0.6cm
{
 {\elevenbf        \vglue 10pt
               A FERMION MASS MATRIX ANSATZ\\
               \vglue 3pt
               FOR THE FLIPPED SU(5) MODEL\footnote{
\ninerm\baselineskip=11pt Invited talk presented at {\it Recent Advances
in the Superworld },
 Houston Advanced Research Center (HARC), The Woodlands, Texas, 14-16
April 1993.} \\}
\vglue 5pt

\vglue 1.0cm
{\tenrm George K. Leontaris \\}
\baselineskip=13pt
{\tenit Physics Department, Ioannina University \\}
\baselineskip=12pt
{\tenit Ioannina, GR-451 10, Greece\\}}

\vglue 0.8cm
{\tenrm ABSTRACT}
\end{center}

A fermion mass matrix ansatz is proposed in the context of Grand Unified
Supersymmetric Theories (GUTs).  The fermion mass matrices are
 evolved down to the electroweak scale by solving the renormalization
group equations for the gauge and Yukawa couplings. Eight
inputs are introduced at the GUT scale to predict the 13
arbitrary parameters of the Standard model. The constraints
imposed by the charged fermion data are used to make
predictions in the neutrino sector. In particular, the neutrino mass matrix
is worked out in the case of the flipped SU(5) model and it is found that the
 {\it generalized } see-saw mechanism which occurs naturally
 in this model can provide  a solution to the solar
neutrino puzzle and give a sufficiently large $\nu_{\tau}$
 mass to contribute as a hot dark matter component as indicated by
the recent COBE data.


\vglue 0.6cm
{\elevenbf\noindent 1. Introduction}
\vglue 0.2cm
\baselineskip=14pt
\elevenrm
Although the Standard Model of Strong and Electroweak interactions explains
all known experimental data, it is now widely accepted that it is unlikely
to be the fundamental theory of nature.

In a fundamental theory the various experimentally measurable parameters
 should be calculable only from few inputs which could be specified in
 terms of the basic principles of the theory. For example, in String
 theories\cite{gsw} one expects that all
masses and mixing angles can be determined only from one input, namely the
String gauge coupling $g_{String}$ at the String Unification scale
$M_{String}\sim O(g_{String}\times M_{Pl})$. However, there is a huge
number of
String derived models and it is rather impossible to find the unique model
among
them which might lead to the correct - experimentally tested - predictions at
low energies. Although the previously mentioned arbitrariness puts a great
obstacle in string model builders, which is not likely to be solved in the near
future, we are by now convinced that there is no real rival to string theory.
{}From the phenomenologist's point of view, the alternative way is to attempt
to
describe the low energy theory, using the experience from string model building
and all the possible information that one can extract out of it.

 We know already, that there is a number of constraints which should be
respected
 on our way from $M_{String}$ down to $m_W$. Here there are listed few of them:

$\bullet $  Recent calculations taking into account threshold corrections from
String massive states\cite{Kap,Ant,Nan,Ross,LT1}, show that the String
Unification scale is relatively high, and close to the Plank scale. On the
contrary, renormalization group calculations have indicated that minimal
supersymmetric Grand Unified Theories (SUSY-GUTs)  are in agreement with the
precision LEP data when the SUSY-GUT scale $M_G$ is taken close to $M_G\approx
10^{16}GeV$\cite{lang}. Thus, the minimal supersymmetric standard model(MSSM)
cannot probably be derived directly at the string scale; the above descrepancy
between the two scales, would rather suggest that MSSM should be obtained
through the spontaneous breaking of some intermediate GUT-like
 gauge group which breaks at the scale $M_G$.

$\bullet $ As long as string model building is based on $k=1$ level
 of Kac-Moody algebras,
it was soon realised that ordinary GUTs (like the SU(5) model), could not be
suitable candidates for a string derived model. The reason is that ordinary
SU(5) needs necessarily Higgs fields in the adjoint in order to break down to
the MSSM, while such representations are not available in $k=1$ constractions.

\vglue 0.5cm

The flipped SU(5) model\cite{barr,aehn,lnz}, was a first attempt to overcome
this
difficulty, and has been proposed as a candidate some time ago\cite{aehn},
 in the
context of the free fermionic formulation of the four-dimensional
 superstring. It was
subsequently discovered\cite{al}, that the GUT symmetry $SU(4)\times
O(4)$\cite{al}, based on the Pati-Salam model\cite{ps}, can also break down
to the standard gauge group without using higgs fields in the adjoint, and
therefore could also serve as a candidate Superstring model.
We should mention that it is also possible to obtain directly the Standard
model gauge group from the string\cite{fny}, but the
proton decay problem\cite{proton} as well as the high string unification
mentioned above, seem to be naturally solved in the presence
of an intermediate
gauge group. Moreover, the Grand Unified models inspired
from strings\cite{aehn,al}
offer natural mechanisms of suppressing the neutrino
masses\cite{neut1,eln}, explain
the baryon asymmetry\cite{basym} of the universe etc.

In view of the above theoretical and phenomenological constraints,
 in my opinon,
the GUT model building issue should be reconsidered and worked out
 systematically.  In the present talk, I will  propose a
specific  fermion mass matrix ansatz\cite{lv}, which was inspired
from phenomenological
calculations  on the flipped string SU(5). The analysis of the charged
 fermion mass spectrum which is given in section 2, has been
done on  completely general grounds, and can apply to any SUSY-GUT model.
Furthermore, in view of the revived interest\cite{dhr,shafi1,ggl,ram}
of the fermion mass problem in GUTs, this work  contributes  also to
the search\cite{lectures} for all possible mass matrices with
the maximum number of zero
entries at the Unification Scale which give the correct predictions at $m_W$.
The discussion on neutrino mass matrix is given in section 3, and applies
mainly to the particular string constructions\cite{aehn,al,fny}.

\vglue 0.6cm
{\elevenbf\noindent 2. Structure of Fermion Mass Matrices at
$M_{GUT}$}
 \vglue 0.4cm
The texture of the quark and lepton mass matrices  has the following
form at the GUT scale
\begin{eqnarray}
M_U&=&\left(\begin{array}{ccc}0&0&x\\0&y&z\\x&z&1\end{array} \right)
\lambda _{top}(t_0) {\upsilon
 \over\sqrt{2}}sin\beta, \label{eq:upq}\\
M_D&=&\left(\begin{array}{ccc}
0&a e^{i\phi}&0\\a e^{-i\phi}&b&0\\0&0&1\end{array} \right)
\lambda _b(t_0){\upsilon
\over\sqrt{2}} cos\beta \label{eq:downq}\\
M_E&=&\left(\begin{array}{ccc}
 0&a e^{i\phi}&0\\a e^{-i\phi}&-3b&0\\0&0&1\end{array} \right)
\lambda _{\tau}(t_0){\upsilon
\over\sqrt{2}} cos\beta \label{eq:elec}\\
M_{\nu_{Dirac}}&=&M_U\label{eq:neutd}
 \end{eqnarray}
with $tan\beta \equiv {<\bar h>
\over <h>}$, $\lambda _b(t_0)=\lambda _{\tau}(t_0)$, and $\upsilon =246 GeV$.
We have taken the up-quark matrix to be symmetric, and the down quark matrix
to be hermitian.
The Dirac neutrino mass matrix, has been taken to be identical to
the up-quark mass matrix
at the GUT scale, since both arise from the same Yukawa term in
these theories\cite{aehn,al}.
 Furthermore, we have chosen to relate the charged lepton mass
 matrix at the GUT scale with
the down quark mass matrix.
 As a matter of fact, the $m_D$ and $m_E$ matrix elements are not
necessarily related in the flipped model. However, our choice
minimizes the number of arbitrary parameters and moreover, it leads
to definite predictions in the neutrino sector.

Let us discuss first  the charged fermion mass matrices.
Our ansatz has a total of five
zeros in  the quark sector (the leptonic sector is directly related to them);
 two zeros for
the up and three for the down quark mass matrix(zeros in symmetric entries
are counted only
once). These zeros reduce the number of arbitrary parameters at the GUT-scale
 to eight, namely
$x,y,z,a,b,\phi,\lambda _b(t_0)$, and $\lambda _{top}(t_0)$. These non-zero
 entries should serve
to determine the thirteen arbitary parameters of the standard model, i.e.,
 nine quark and
lepton masses, three mixing angles and the phase of the
Cabbibo-Kobayashi-Maskawa (CKM)
matrix. Thus, as far as the charged fermion sector is concerned,
we end up with five
predictions (we will discuss the predictions in the neutrino sector
in the next section.).
 We may reduce the number of arbitary parameters
by one, if we impose more structure in the up-quark mass matrix.
We may for example relate
the $(13)$, $(23)$ and $(22)$ entries, as follows
\begin{eqnarray}
y &= & n z^2  \\
 x& = & (n-1) z^2
\end{eqnarray}

where $n$ can be a number in the range $n\sim (3,10)$. Although this
 structure is imposed by
hand and not actually necessary\cite{lv}, the results are more presentable
  and calculations
more easy to handle.


  In order to find the  structure of the mass matrices at
the low energy scale and calculate the mass eigenstates as well as
the mixing matrices and compare them  with the experimental data, we
need to evolve them down to $m_W$, using the renormalization group
equations.
The renormalization group
equations for the Yukawa couplings at the one-loop level, read

\begin{eqnarray}
16\pi^2 \frac{d}{dt} \lambda_U&=& \left( I\cdot
Tr [3 \lambda_U\lambda_U^\dagger ]  +
3 \lambda_U \lambda_U^\dagger +\lambda_D \lambda_D^\dagger
-I\cdot G_U\right) \lambda_U, \label{eq:rge1}
\\
16\pi^2 \frac{d}{dt} \lambda_N&=& \left( I\cdot Tr [
\lambda_U \lambda_U^\dagger ]  + \lambda_E \lambda_E^\dagger -I
\cdot G_N\right) \lambda_N, \label{eq:rge2}
\\
16\pi^2 \frac{d}{dt} \lambda_D&=& \left( I\cdot Tr
[3 \lambda_D\lambda_D^\dagger +
\lambda_E \lambda_E^\dagger ]  + 3 \lambda_D \lambda_D^\dagger
 +\lambda_U \lambda_U^\dagger
-I \cdot  G_D\right) \lambda_D, \label{eq:rge3}
\\
16\pi^2 \frac{d}{dt} \lambda_E&=& \left( I\cdot
Tr [ \lambda_E\lambda_E^\dagger +3
\lambda_D  \lambda_D^\dagger ]  +
3 \lambda_E \lambda_E^\dagger -I \cdot G_E\right) \lambda_E,
\label{eq:rge4}
\end{eqnarray}
where $\lambda_\alpha$, $\alpha=U,N,D,E$, represent the $3$x$3$
Yukawa matrices
which are defined in terms of the mass matrices given in
Eq.(\ref{eq:upq}-\ref{eq:neutd}), and $I$ is the $3$x$3$ identity matrix.
We have neglected one-loop corrections proportional to $\lambda_N^2$.
$t\equiv\ln(\mu/\mu_0)$, $\mu$ is the scale at which the couplings are to be
determined and $\mu_0$ is the reference scale, in our case the GUT scale. The
gauge contributions are given by
\begin{eqnarray}
G_\alpha&=&\sum_{i=1}^3 c_\alpha^i g_i^2(t),\\
g_i^2(t)&=&\frac{g_i^2(t_0)}{1- \frac{b_i}{8\pi^2} g_i^2(t_0)(t-t_0)}.
\end{eqnarray}
The $g_i$ are the three gauge coupling constants of the Standard Model and
$b_i$
are the corresponding supersymmetric beta functions. The coefficients
$c_\alpha^i$ are given by
\begin{eqnarray}
\{c_U^i \}_{i=1,2,3} &=& \left\{ \frac{13}{15},3,\frac{16}{3} \right\}, \qquad
\{c_D^i \}_{i=1,2,3} = \left\{\frac{7}{15},3,\frac{16}{3} \right\}, \\
\{c_E^i \}_{i=1,2,3} &=& \left\{ \frac{9}{5},3,0\right\}, \qquad \quad
\{c_N^i \}_{i=1,2,3} = \left\{ \frac{3}{5},3,0\right\}.
\end{eqnarray}
In what follows, we will assume that $\mu_0\equiv M_G\approx 10^{16}GeV$
and $a_i(t_0)\equiv
 \frac{g_i^2(t_0)}{4\pi}\approx \frac{1}{25}$.
In order to evolve the equations(Eqs) (\ref{eq:rge1}-\ref{eq:rge4}) down
to low energies, we also
need to do some plausible approximations. First, we find it convenient
to diagonalize the up
quark mass matrix at the GUT scale and obtain the eigenstates
\begin{eqnarray}
m_1(M_{GUT})\approx -n(n-1)ptan^2\theta sin^2\theta \nonumber\\
m_2(M_{GUT})\approx (n-1)ptan^2\theta\label{eq:Gmass}\\
m_3(M_{GUT})\approx \frac{p}{cos^2\theta}\nonumber
\end{eqnarray}

with diagonalizing matrix
\begin{eqnarray}
{\it K}&=&\left(\begin{array}{ccc}

\frac{1}{{\it D_1}}&\frac{-sin\theta}{{\it D_2}}&\frac{(n-1)sin^2\theta}
{{\it D_3}}\\
\frac{sin2\theta}{2{\it D_1}}&\frac{1}{{\it D_2}}&\frac{sin2\theta}
{{\it D_3}}\\
\frac{nsin\theta}{{\it D_1}}&\frac{nsin\theta}{{\it D_2}}&
\frac{1-nsin^2\theta}{{\it D_3}}
\end{array} \right) \label{eq:upk}
\end{eqnarray}
where $p=\lambda_t(t_0){\upsilon \over\sqrt{2}}sin\beta $, ${\it D_1}
\approx \sqrt{
1+sin^2\theta cos^2\theta}$, ${\it D_2}\approx \sqrt{1+sin^2\theta }$, and
${\it D_3}\approx \sqrt{1-(2n-1)sin^2\theta }$. In this case all other
 mass matrices are
also rotated by the same similarity transformation, thus
\begin{eqnarray}
\lambda_{\alpha}&\rightarrow &{\tilde\lambda}_{\alpha}=
 K^\dagger\lambda_{\alpha}K,\alpha
=D,E,N \end{eqnarray}
 Now assuming that the $\lambda_t$ coupling is much bigger
than all other fermion Yukawa couplings, we may ignore the contributions
of the latter in
the RHS of the RGEs in Eqs(\ref{eq:rge1}-\ref{eq:rge4}).
 In this case all differencial equations
reduce to a simple uncoupled form. Thus the top-Yukawa coupling
differential equation, for
example, may be cast in the form
\begin{eqnarray}
16\pi^2 \frac{d}{dt}
\tilde\lambda_{top}&=&\tilde\lambda_{top}(6{\tilde\lambda}_{top}
^2-G_U(t))\label{eq:topeq} \end{eqnarray}
with the solution\cite{Lopez,ggl}
\begin{eqnarray}
\tilde \lambda_{top}&=&\tilde \lambda_{top}(t_0)\xi^6\gamma_U(t)\
 \label{eq:ltop}
\end{eqnarray}
where
\begin{eqnarray}
\gamma_U(t)&=& \prod_{j=1}^3 \left(1- \frac{b_{j,0}\alpha_{j,0}(t-t_0)}{2\pi}
\right)^{c_\alpha^j/2b_j},\\
\xi&=& \left( 1-\frac{6}{8\pi^2}\lambda_{top}(t_0)
\int_{t_0}^{t} \gamma_U^2(t)\,dt \right)^{-1/12}\label{eq:ksi}.
\end{eqnarray}
Thus the GUT up quark mass eigenstates given in Eq(\ref{eq:Gmass})
 renormalized down to their
mass scale, are given by
\begin{eqnarray}
m_u \approx \gamma_U\xi^3 \eta_u n(n-1)ptan^2\theta sin^2\theta \nonumber\\
m_c \approx \gamma_U\xi^3\eta_c (n-1)ptan^2\theta\label{eq: upmass}\\
m_t \approx \gamma_U\xi^6 \frac{p}{cos^2\theta}\nonumber
\end{eqnarray}
where $\eta_u$ , $n_c$ are taking into account the renormalization effects
 from
the $m_t$-scale down to the mass of the corresponding quark.
We may combine the equations in
Eq(\ref{eq: upmass}) and give predictions for the top quark mass and the
 mixing
angle $\theta $
in terms of the low energy up and charm quark masses. We obtain
 \begin{eqnarray}
m_t \approx \xi^3\frac{n}{n-1}\frac{m_c^2}{m_u}\frac{\eta_u}
{\eta_c^2}\label{eq: tpred}\\
sin\theta =\sqrt{\frac{m_u}{nm_c}}\label{eq:thpred}
\end{eqnarray}
In the basis where the up-quark mass matrix is diagonal,
the renormalized down-quark and
charged lepton mass matrices are given by
\begin{eqnarray}
m_D^{ren} &\approx & \gamma_D I_{\xi}{\it K^{\dagger}}m_D{\it K}
\label{eq:rdown}\\
m_E^{ren} &\approx & \gamma_E {\it K^{\dagger}}m_E{\it K} \label{eq:relec}
\end{eqnarray}
where $I_{\xi}=Diagonal(1,1,\xi)$, and $m_{D,E}$, are given
in Eqs(\ref{eq:downq}) and
(\ref{eq:elec}).
 In order to make definite predictions in the fermion mass sector,
 we choose
the lepton masses as inputs and express the arbitrary parameters
$a,b,\lambda_{\tau}(\equiv
\lambda_b) $ in $m_E$ in terms of $m_e,m_{\mu}$, and  $m_{\tau}$.
Substituting into $m_D$ and
diagonalizing $m_D^{ren}$, we obtain\cite{lv}
\begin{eqnarray}
m_d\approx  6.3\times (\frac{\eta_d}{2})MeV \\
m_s\approx 153 \times (\frac{\eta_s}{2})MeV \\
m_b \approx \eta_b \frac{\gamma_D}{\gamma_E}\xi m_{\tau}\label{eq:bottom}
\end{eqnarray}
Since $\eta_d\approx \eta_s\approx 2$, the predictions for the light quarks
$m_{d,s}$ are
within the expected range\cite{qcd}. Now, in order to make a prediction
 for the bottom quark, we
need to know the value of $\xi$, but the latter depends on the top quark
coupling at the GUT
scale  as well as on the top-mass, through Eq.(\ref{eq:ksi}). We can use,
 however,
Eq.(\ref{eq:bottom}), to predict the range of the top mass.
 Thus, using the available limits
on the bottom mass, $4.15 \le m_b\le 4.35 GeV$, and $\eta_b\approx 1.4$,
we can obtain the following range for $m_{top}$
\begin{eqnarray}
125GeV\le m_{top}  \le (165-170) GeV\label{eq:toprange}
\end{eqnarray}
The CKM - matrix can be found by diagonalizing $m_D^{ren}$
in Eq(\ref{eq:rdown}). For all the
above range of the $m_{top}$ mass, it is always possible to find a set
 of the input
parameters, which give CKM-mixing angles within the experimentally
 acceptable ranges. Its
analytic form is rather complicated. As an example, we give the CKM-entries
 for the specific
case where, $m_t=135GeV$, $tan\beta =1.1$,and $\phi =\frac{\pi}{4}$.
In this case $m_b$ is
predicted to be $4.33GeV$, while we obtain
\begin{eqnarray}
{\mid V_{CKM} \mid}_{ij}&=&\left(\begin{array}{ccc}
0.9754&0.22&0.0032\\0.22&0.9750&0.038\\0.01&0.037&0.9993\end{array} \right)
 \label{eq:ckm}
\end{eqnarray}
\vglue 0.6cm
{\elevenbf\noindent 2. Neutrino Masses}
 \vglue 0.4cm
There is a lot of experimental evidence today, that the neutrinos
have non-zero masses.
For example, recent data from solar neutrino experiments\cite{H1}
 show that the
deficiency of solar neutrino flux, i.e. the discrepancy
between theoretical estimates and the experiment, is
naturally explained if the $\nu _e$ neutrino oscillates to
another species during its flight to the earth.
Furthermore, the COBE
measurement \cite{Smooth} of the large scale
microwave background anisotropy, might be
explained \cite{SShafi} if one assumes an
admixture of cold ($ \sim 75\% $) plus hot ($\sim25\% $)
dark-matter. It is hopefully expected that some
 neutrino (most likely $\nu _{\tau }$)
  may be the natural candidate of the hot dark
matter component.

Here we would like to address the question of
neutrino masses in GUT models
which arise \cite{aehn,al}in the free fermionic construction of
four dimensional strings.  Taking into
account renormalization effects , it was recently shown\cite{eln,lv} that the
general see-saw mechanism which occurs naturally in the flipped
model, turns out to be consistent with the recent solar neutrino
data, while on the other hand suggests that CHOROUS and NOMAD
experiments at CERN may have a good chance of observing
$\nu_{\mu}\longleftrightarrow  \nu_{\tau }$ oscillations\footnote{
\ninerm\baselineskip=11pt for neutrinos in convensional GUTs
 see\cite{various}}

The various tree-level superpotential mass terms which contribute to
the neutrino mass matrix in the flipped $SU(5)$ model are the
following:
\begin{eqnarray}
\lambda _{ij}^{u}F^{i}\bar f^{j}\bar h +\lambda
_{ij}^{\phi \nu ^{c}}F^{i}\bar H\phi ^{j}+ \lambda _{ij}^{\phi}\phi ^0\phi
^{i}\phi ^{j} \label{eq:potential}
\end{eqnarray}
where in the above terms $F^{i},\bar f^{j}$ are the $ {10},\bar 5$
matter SU(5) fields while $\bar H,\bar h, h$ are the $\bar {10},\bar
5, 5$ Higgs representations and $\phi^{i}$ are neutral $SU(5)\times
U(1)$  singlets. The Higgs field $\bar H$ gets a vacuum expectation
value(v.e.v.) of the order of the SU(5) breaking scale ($\sim
10^{16}GeV$), $\bar h, h$ contain the standard higgs doublets while
$\phi ^0$ acquires a v.e.v., most preferably at the electoweak scale.
The neutrino mass matrix may also receive significant contributions
from other sources. Of crusial importance, are the
non-renormalizable contributions\cite{enkl} which may give a direct
$M_{\nu ^c\nu ^c}=M^{rad}$ contribution which is absent in the
tree-level potential. Then, the general $9\times 9$ neutrino mass
matrix in the basis $(\nu _i,\nu _i^c,\phi _i)$, may be written as
follows:

\begin{eqnarray}
m_{\nu }=
\left(\matrix{0&m_U&0\cr
m_U&M^{rad}&M_{\nu ^{c},\phi }\cr
0&M_{\nu ^{c},\phi }&\mu _{\phi }\cr}\right) \label{eq:SeeSaw}
\end{eqnarray}
where it is understood that all entries in Eq.(\ref{eq:SeeSaw}) represent
 $3\times
3$ matrices. The above neutrino matrix is different from that of
standard see-saw matrix-form, since now there are three neutral
 $SU(5)\times U(1)$ singlets
involved, one for each family.

It is clear that the matrix (\ref{eq:SeeSaw}) depends on a relatively
large number of
parameters and a reliable estimate of the light neutrino masses and the
mixing angles  is a rather complicated task. We are going to use however
 our knowledge of the rest of the fermion spectrum to reduce
sufficiently the number of parameters involved. Firstly, due to the GUT
relation $m_U(M_{GUT})=m_{\nu _D}(M_{GUT})$, we can deduce the form of
$m_{\nu _D}(M_{GUT})$, at the GUT scale in terms of the up-quark masses.
The heavy majorana $3\times 3$ matrix $M^{rad}$, depends on the kind of the
specific generating mechanism. Here\cite{lv} we take it to be proportional
 to the
down quark-matrix at the GUT scale:
\begin{eqnarray}
M^{rad}=\Lambda ^{rad} m_D(M_{GUT})\label{eq:mrad}
\end{eqnarray}
 The
$M_{\nu ^{c},\phi }$ and $\mu _{\phi }$  $3\times 3$ submatrices are
also model dependent. In most of the string models however, there is
only one entry at the trilinear superpotential in the matrix $M_{\nu
^{c},\phi }$, which is of the order $M_{GUT}$. Other terms, if any,
usually arise from high order non-renormalizable terms. We will
assume in this work  only the existence of the trilinear term,
since higher order ones will be comparable to  $M^{rad}$  and are
not going to change our predictions. In particular we will take
$M_{\nu ^{c},\phi }\sim Diagonal(M,0,0)$, and $\mu _{\phi }\sim
Diagonal(\mu,0,0)$, with $\mu << M\sim M_{GUT}$, thus we will treat
 Eq.(\ref{eq:SeeSaw}) as a $7\times 7$ matrix.

To obtain the neutrino spectrum and lepton mixing, we must
introduce values for the two additional papameters $M, \Lambda
^{rad} $ of the neutrino mass matrix in Eq.(\ref{eq:SeeSaw}).
 We assume naturally $M=<\bar H>\approx 10^{16}GeV$.
The neutrino mass eigenvalues can now be predicted in terms
of the scale quantity
$\Lambda ^{rad}$ . Thus they can be written as\cite{lv}
\begin{eqnarray}
m_{\nu _e}\approx 0,
m_{\nu _{\mu}}={\Lambda ^{\mu}\over \Lambda
^{rad}}\times 10^{-2}eV,
m_{\nu _{\tau}}={\Lambda ^{\tau}\over \Lambda
^{rad}}\times 10eV\label{eq:eigeneut}
\end{eqnarray}
For $m_t\approx 130GeV$ we get $\Lambda ^{\mu} \approx .80\times
10^{12}$ and  $\Lambda ^{\tau} \approx 1.85\times 10^{12}$.

For the oscillation probabilities, we find\cite{lv}
\begin{eqnarray}
P(\nu _e\leftrightarrow \nu _\mu ) \approx 3.1\times 10^{-2}
sin^2({\pi {L\over l_{12}}})\label{eq:osc12}
\end{eqnarray}

 \begin{eqnarray}
 P(\nu _{\tau}\rightarrow \nu _{\mu }) \approx
4.0\times 10^{-3} sin^2({\pi {L\over l_{13}}})\label{eq:osc23}
\end{eqnarray}
\begin{eqnarray}
 P(\nu _e\rightarrow \nu _{\tau }) \approx
4.0\times 10^{-5} sin^2({\pi {L\over l_{13}}})\label{eq:osc13}
\end{eqnarray}
where L is the source--detector distance and
\begin{eqnarray}
l_{ij}={4\pi E_{\nu}\over |m_i^2-m_j^2|}\label{eq:oscl}
\end{eqnarray}

We can determine the range of
$\Lambda ^{rad}$, using the available solar neutrino data
(see for example ref\cite{petcov} for a systematic analysis of the
allowed regions using all
available data)
\begin{eqnarray}
5.0\times 10^{-3}\le sin^22\theta_{ij}\le 1.6\times
10^{-2}\label{eq:data1}
\end{eqnarray}
\begin{eqnarray}
0.32\times 10^{-5}(eV)^2\le \delta m_{ij}^2\le 1.2\times
10^{-5}(eV)^2\label{eq:data2}
\end{eqnarray}
 Our result in Eq(\ref{eq:osc12})
is a bit outside the above range but the mass constraint can be
easily satisfied by choosing $\Lambda ^{rad}$ in the range
\begin{eqnarray}
.7\times 10^{12}\le \Lambda ^{rad}\le 7\times 10^{12}\label{eq:lamrad}
\end{eqnarray}
 Our neutrino masses can also  easily be made to fall into the range
of the Frejus atmospheric neutrinos
\begin{eqnarray}
10^{-3}(eV)^2\le \delta m_{ij}^2\le 10^{-2}(eV)^2
\end{eqnarray}
but our mixing is much too small. Our results are also consistent
with the data on $\nu _{\mu } \leftrightarrow \nu _{\tau } $
 oscillations\cite{mutau}
\begin{eqnarray}
 sin^22\theta_{\mu \tau}\le 4.\times 10^{-3},\delta m_{\nu _{\mu}
\nu _{\tau}}^2\ge 50(eV)^2\label{eq:mtosc}
\end{eqnarray}
Our results however cannot be made to fall on the $sin^22\theta $
$vs$  $\delta m^2$ of the $BNL$ $\nu _{\mu } \leftrightarrow \nu _e $
oscillation results\cite{boro}.

Moreover, it is always possible to obtain $m_{\nu _{\tau }}\approx
(few \sim 20)eV$, hence one can obtain simultaneously the
cosmological HOT-dark matter component,
 in agreement with the interpretation of the COBE data.
 Indeed
an upper limit on the $\nu _{\tau }$ mass can be obtained from
the formula
\begin{eqnarray}
7.5\times 10^{-2}\le \Omega _{\nu }h^2\le 0.3\label{eq:ntau1}
\end{eqnarray}
Translating this into a constraint on $m_{\nu {_\tau}}$, arising
from the relation $m_{\nu_{\tau }}\approx \Omega _{\nu }h^291.5eV$
where $h=.5\sim 1.0$ is the Hubble parameter, one gets the range
\begin{eqnarray}
6.8\le m_{\nu _{\tau }}\le 27eV\label{eq:ntau1}
\end{eqnarray}
 which can be easily achieved with the above range of $\Lambda
^{rad}$.
 \vglue 0.2cm
{\elevenbf Acknowledgements} {\elevenit I would like to thank the
Astroparticle Physics Group,
the organizing committee of the workshop, and in particular
 D.V. Nanopoulos, J. Lopez, I.
Giannakis, K. Yuan and G. Kyriazis for kind hospitality}.
\vglue 0.3cm
{\elevenbf\noindent 3. References \hfil}  \vglue 0.4cm

\end{document}